\documentclass[a4paper,11pt]{article}
\usepackage{graphicx}
\usepackage{amsfonts}
\usepackage{amssymb}
\begin{document}
\begin{titlepage}
\title{Approximation of the naive black hole degeneracy}
\maketitle
\begin {center}
\author{\large Youngsub Yoon }

\vskip 1cm

{
\it
Department of Physics and Astronomy\\
Seoul National University, Seoul 151-747, Korea
}
\end{center}

\begin{abstract}
In 1996, Rovelli suggested a connection between black hole entropy and the area spectrum. Using this formalism and a theorem we prove in this paper, we briefly show the procedure to calculate the quantum corrections to the Bekenstein-Hawking entropy. One can do this by two steps.
First, one can calculate the ``naive'' black hole degeneracy without the projection constraint (in case of the $U(1)$ symmetry reduced framework) or the $SU(2)$ invariant subspace constraint (in case of the fully $SU(2)$ framework). Second, then one can impose the projection constraint or the $SU(2)$ invariant subspace constraint, obtaining logarithmic corrections to the Bekenstein-Hawking entropy. In this paper, we focus on the first step and show that we obtain infinite relations between the area spectrum and the naive black hole degeneracy. Promoting the naive black hole degeneracy into its approximation, we obtain the full solution to the infinite relations. 
\end{abstract}

\end{titlepage}
\pagebreak

\section {Introduction}
It is well known that the entropy of a black hole is given by the quarter of its area (i.e. $A/4$), regardless of the type of black hole considered \cite{Bekenstein, Hawking}. However, as it is so only in the leading order, many have calculated the corrections to it \cite{corrections, Meissner}.

In this paper, we will consider the connection between the black hole entropy and the area spectrum suggested by Rovelli in 1996 \cite{Rovelli} as loop quantum gravity predicts that the area spectrum is quantized \cite{Discreteness, General, Ashtekar}. To apply this connection, we will consider the formula proposed by Domagala, Lewandowski and Meissner which can check whether Bekenstein-Hawking entropy is consistent with a given area spectrum \cite{Meissner, Domagala}. Stepping further, we will use the mathematics of ``compositions,'' to prove a theorem that shows that Bekenstein-Hawking entropy is reproduced, if their formula is satisfied. Then, by basing on this formalism, we calculate the ``naive'' degeneracy of black hole. We call it ``naive'' as we calculated it without the consideration of the projection constraint (in case of the $U(1)$ symmetry reduced framework) or the SU(2) invariant subspace constraint (in case of the fully $SU(2)$ framework). During the process, we obtain infinite relations between the area spectrum and the naive black hole degeneracy. Then, we ``continutize'' or ``approximate'' the naive black hole degeneracy to the smooth function of area and obtain the full solution to the infinite relations. This is the main result and objective of this paper.

The organization of this paper is as follows. In section 2, we introduce the relation between black hole entropy and the area spectrum proposed by Rovelli. In section 3, we introduce Domagala-Lewandowksi-Meissner formula. In section 4, we introduce the mathematics of ``compositions.'' In section 5, we prove that the Bekenstein-Hawking entropy is reproduced, if the area spectrum satisfies Domagala-Lewandowski-Meissner formula. In section 6, we apply this formalism to calculate the ``naive'' black hole degeneracy. We will also obtain the infinite relations just advertised. In section 7, we will obtain the full solution to the infinite relations. In section 8, we show, as an example, how one can obtain logarithmic correction; the result of this section is nothing new. We consider the $U(1)$ symmetry reduced framework and show how the projection constraint $\sum_i m_i=0$ yields the logarithmic corrections. In section 9, we conclude our paper.

\section {Black hole entropy and the area spectrum}
According to loop quantum gravity, the eigenvalues of the area operator are discrete. Let's say that we have the following area eigenvalues, or the unit areas:

\begin{equation}
A_1, A_2, A_3, A_4, A_5, A_6....
\end{equation}
Here, we used the notation that the $i$th unit area is $A_i$. Then, a generic area should be partial sum of them, including the case in which the same area eigenvalues are repeated in the sum. In other words, a generic area has many partitions, each of which must be $A_i$ for some $i$.

Given this, an interesting proposal was made by Rovelli \cite{Rovelli}. As black hole entropy is given by $A/4$, the degeneracy of black hole is given by $e^{A/4}$. Rovelli proposed that the black hole degeneracy is obtained by counting the number of ways in which the area of black hole can be expressed as the sum of unit areas. In other words, $N(A)$ the degeneracy of the black hole with area $A$ is given by following.

\begin{equation}
N(A):=\left\{(i_{1},i_{2},i_{3} \cdots), \sum_{x} A_{i_x}= A\right\}\label{NA}
\end{equation}

Here, I want to note an important point. For the parenthesis in the above formula $(\cdots,a,\cdots,b,\cdots)$ should be regarded different from $(\cdots,b,\cdots,a,\cdots)$. In other words, the order in the summation is important.

\section {Domagala-Lewandowski-Meissner trick}
This section closely follows \cite{Tanaka} which explains Domagala-Lewandowski-Meissner trick in an easier way. To understand their formula which gives a necessary condition for the black hole entropy to be $A/4$, we reconsider the ``simplified area spectrum'' or``isolated horizon'' as follows \cite{isolated horizon}. In this case, $A_i=8 \pi \gamma \sqrt{j_i(j_i+1)}$. Then, (\ref{NA}) becomes the following.
\begin{equation}
N(A):=\left\{(j_{1}, \cdots,j_{n})|0\neq j_{i}\in \frac{\mathbb{N}}{2}, \sum_{i}\sqrt{j_{i}(j_{i}+1)}=\frac{A}{8\pi\gamma}\right\}
\end{equation}

We derive a recursion relation to obtain the value of $N(A)$. When we consider $(j_{1}, \cdots,j_{n})\in N(A-a_{1/2})$ we obtain $(j_{1}, \cdots,j_{n},\frac{1}{2})\in N(A)$, where $a_{1/2}$ is the minimum area where only one $j=1/2$ edge contributes to the area eigenvalue. i.e., $a_{1/2}=8\pi\gamma\sqrt{\frac{1}{2}(\frac{1}{2}+1)}=4\pi\gamma\sqrt{3}$. Likewise, for any eigenvalue $a_{j_{x}}(0< a_{j_{x}}\leq A)$ of the area operator, we have

\begin{equation}
(j_{1}, \cdots,j_{n})\in N(A-a_{j_{x}})\Longrightarrow (j_{1}, \cdots,j_{n},j_{x})\in N(A).
\end{equation}

Then, important point is that if we consider all $0< a_{j_{x}}\leq A$ and $(j_{1}, \cdots,j_{n})\in N(A-a_{j_{x}})$, $ (j_{1}, \cdots,j_{n},j_{x})$ form the entire set $N(A)$. Thus, we obtain

\begin{equation}
N(A)=\sum_{j} N(A-8\pi\gamma\sqrt{j(j+1)})
\end{equation}

By plugging $N(A)=\exp(A/4)$, one can determine whether the above formula satisfies Bekenstein-Hawking entropy formula. If the Bekenstein-Hawking entropy is satisfied, from the above formula, we have \cite{Domagala,Meissner}:

\begin{equation}
1=\sum_{j} \exp(-8\pi\gamma\sqrt{j(j+1)}/4)\label{section6ashtekar}
\end{equation}

In other words,
\begin{equation}
1=\sum_{i} e^{-A_i/4}\label{DLM}
\end{equation}

\section {Compositions}
This section closely follows \cite{Compositions, MacMahon}. A composition is an integer partition in which order is taken into account. For example, there are eight compositions of $4$: $4, 3+1, 1+3, 2+2, 2+1+1, 1+2+1, 1+1+2$ and $1+1+1+1$. $c(n)$ denotes the number of compositions of n, and $c_m(n)$ is the number of compositions into exactly $m$ parts. For example: $c(4)=8$, $c_3(4)=3$.

It is easy to understand that $c_m(n)$ is given by the coefficient of $x^n$ in the expansion of
\begin{equation}
(x+x^2+x^3+\cdots)^m
\end{equation}
for writing the function as a product of $m$ factors and performing the multiplication by picking out the terms $x^{p_1}, x^{p_2},\cdots,x^{p_m}$, where
\begin{equation}
p_1+p_2+\cdots+p_m=n
\end{equation}
in succession, we obtain for this particular selection the term $x^{p_1+p_2+\cdots+p_m}$ of the product, where $(p_1,p_2,\cdots,p_m)$ is one composition of $n$ into exactly $m$ parts.

In other words,
\begin{equation}
(x+x^2+x^3+\cdots)^m=\sum_{n=1}^{\infty} c_m(n) x^n
\end{equation}

Then, as
\begin{equation}
c(n)=\sum_{m=1}^{\infty} c_m(n)
\end{equation}
we have
\begin{equation}
\sum_{m=1}^{\infty}(x+x^2+x^3+\cdots)^m=\sum_{n=1}^{\infty} c(n) x^n
\end{equation}

Now, we can explicitly calculate $c(n)$. The above formula is equal to:

\begin{equation}
\frac{(x+x^2+x^3+\cdots)}{1-(x+x^2+x^3+\cdots)}=\frac{x}{1-2x}
\end{equation}
Therefore, we obtain $c(n)=2^{n-1}$

\section {Our theorem}
Now, let's apply the lesson from our earlier section to our case, namely, Bekenstein-Hawking entropy. The fact that $\left\{A_i, A_j\right\}$ should be regarded different from $\left\{A_j, A_i\right\}$ suggests that the calculation of black hole entropy has a similar structure to ``compositions'' in which the order is taken into account. Considering this, (\ref{NA}) can be translated into

\begin{equation}
\sum_{m=1}^{\infty}(e^{-s A_1}+e^{-s A_2}+\cdots)^m=\frac{e^{-s A_1}+e^{-s A_2}+\cdots}{1-(e^{-s A_1}+e^{-s A_2}+\cdots)}=\sum_A N(A) e^{-s A}\label{NAE-SA}
\end{equation}
where $s$ is an arbitrary parameter. It is easy to see that the above formula converges for $s$ such that
\begin{equation}
e^{-s A_1}+e^{-s A_2}+\cdots<1 \label{e-sA_1}
\end{equation}
and diverges for $s$ such that
\begin{equation}
e^{-s A_1}+e^{-s A_2}+\cdots\geq1 \label{e-sA_2}
\end{equation}

However, from Domagala-Lewandowski-Meissner formula (\ref{DLM}), we have:
\begin{eqnarray}
e^{-A_1/4}+e^{-A_2/4}+\cdots=1 \label{e-A1/4}
\end{eqnarray}

Therefore, by examining (\ref{e-sA_1}) and (\ref{e-sA_2}), we can see that (\ref{NAE-SA}) converges for $s>\frac{1}{4}$, and diverges for $s\leq\frac{1}{4}$. Given this, if we closely examine the right-hand side of (\ref{NAE-SA}), the only conclusion that we can draw is that (\ref{e-A1/4}) implies
\begin{equation}
N(A)\sim P(A) e^{A/4}
\end{equation}
for large enough $A$, and for $P(A)$ which does not increase or decrease faster than an exponential function.

\section {The naive black hole degeneracy}
Let's focus on the behavior of (\ref{NAE-SA}), when $s$ is slightly bigger than $\frac{1}{4}$. We write:

\begin{equation}
s=\frac{1}{4}+\alpha
\end{equation}

And, let's use following notation, which should be familiar from statistical mechanics.
\begin{equation}
<P(A)>\equiv\sum_i P(A_i) e^{-A_i/4}
\end{equation}

Notice that the above formula is correctly normalized, as $<1>=1$. Then, by Taylor expansion, we have:
\begin{equation}
\sum_i e^{-(1/4+\alpha)A_i}=1-\alpha<A>+\frac{\alpha^2}{2}<A^2>-\frac{\alpha^3}{6}<A^3>+\cdots
\end{equation}

Plugging the above formula to (\ref{NAE-SA}), we obtain:
\begin{equation}
\sum_A N(A) e^{-(1/4+\alpha)A}=\frac{1}{\alpha <A>}+(\frac{<A^2>}{2<A>^2}-1)+(-\frac{1}{3}\frac{<A^3>}{<A>^2}+\frac{1}{4}\frac{<A^2>^2}{<A>^3})\alpha+\cdots\label{crucial}
\end{equation}

Now, let's reexpress the left-hand side of the above formula. Using the following notation,

\begin{equation}
P(A)\equiv N(A)e^{-A/4}\label{PAE-A4}
\end{equation}
and considering the fact that one can approximate summation in terms of integration in the limit $A_{cut}\rightarrow \infty$, we can write:

\begin{eqnarray}
&& \sum_A N(A) e^{-(1/4+\alpha)A}=\lim_{A_{cut}\rightarrow \infty}\{\sum_{A<A_{cut}} N(A) e^{-(1/4+\alpha)A}+ \int_{A_{cut}}^{\infty} N(A)e^{-(1/4+\alpha)A} dA\} \nonumber\\
&& =\lim_{A_{cut}\rightarrow \infty} \lim_{\alpha A_{cut}\ll 1}\{\sum_{A<A_{cut}} P(A) (1-\alpha A +\frac{\alpha^2 A^2}{2}+\cdots)+\int_{A_{cut}}^{\infty} P(A)e^{-\alpha A} dA\}\label{compare}
\end{eqnarray}

This separation of $N(A)$ into the case when $A$ is small and the case when $A$ is big is useful, as when $A$ is too small, the ``fluctuation'' or the ``randomness'' of $N(A)$ is so big that it cannot be approximated by a well-behaving function of $A$. Moreover, it will turn out soon that the last term in the above formula would diverge, if we didn't do the separation and took the whole range of $A$ into the consideration. (i.e. if $A_{cut}$=0) Therefore, the separation is essential. Now, we must compare the above formula with (\ref{crucial}). We easily see the following:
\begin{equation}
\lim_{A_{cut}\rightarrow \infty}\lim_{\alpha A_{cut}\ll 1}\int_{A_{cut}}^{\infty} P(A)e^{-\alpha A} dA=\frac{1}{\alpha <A>}+O(1)+O(\alpha)+\cdots\label{PAE-AA}
\end{equation}
which suggests the following approximation for large $A$:
\begin{equation}
P(A)\approx P_0+\frac{P_1}{A}+\frac{P_2}{A^2}+\cdots\label{Laurent}
\end{equation}
as
\begin{eqnarray}
\lim_{A_{cut}\rightarrow \infty}\lim_{\alpha A_{cut}\ll 1}\int_{A_{cut}}^{\infty} (P_0+\frac{P_1}{A}+\frac{P_2}{A^2}+\cdots)e^{-\alpha A}dA\label{cestbien} \nonumber\\
=\frac{P_0}{\alpha}+O(1)+O(\alpha)+\cdots\label{finally}
\end{eqnarray}

In other words, in order that (\ref{PAE-AA}) and (\ref{finally}) match each other order by order, the terms proportional to the positive powers of $A$ are absent in (\ref{Laurent}). This implies:

\begin{equation}
P_0=\frac{1}{<A>}\label{implication}
\end{equation}

Now, let's explicitly consider the matching for the higher-order terms. If we take a derivative of (\ref{cestbien}) with respect to $\alpha$, we get:

\begin{eqnarray}
\lim_{A_{cut}\rightarrow \infty}\lim_{\alpha A_{cut}\ll 1}&& \int_{A_{cut}}^{\infty} -(P_0 A+P_1+\frac{P_2}{A}+\cdots)e^{-\alpha A}dA \nonumber\\
&& ~~~~=-P_0\frac{e^{-\alpha A_{cut}}}{\alpha^2}-P_0 \frac{A_{cut}}{\alpha}e^{-\alpha A_{cut}}-\frac{P_1}{\alpha}e^{-\alpha A_{cut}}+\cdots \nonumber\\
&& ~~~~=-\frac{P_0}{\alpha^2}-\frac{P_1}{\alpha}+\cdots\label{interpretation}
\end{eqnarray}
where we have Taylor expanded $e^{-\alpha A_{cut}}$ in the last step.

Given this, notice that the above formula must be equal to the following:
\begin{eqnarray}
\lim_{A_{cut}\rightarrow \infty}\lim_{\alpha A_{cut}\ll 1}\int_{A_{cut}}^{\infty} -(P_0 A+P_1+\frac{P_2}{A}+\cdots)e^{-\alpha A}dA = -\frac{P_0}{\alpha^2}+O(1)+O(\alpha)+\cdots
\end{eqnarray}
which is the derivative of (\ref{finally}) with respect to $\alpha$. In other words, the term proportional to $1/\alpha$ is absent in the above formula. This suggests:

\begin{equation}
P_1=0
\end{equation}
Let us give you some interpretations for this result. A non-zero $P_1$ suggests that (\ref{interpretation}) implies the presence of the term $P_1 \ln \alpha$ in (\ref{finally}). However, a term proportional to $\ln \alpha$ is absent in (\ref{PAE-AA}). So, we conclude $P_1=0$.

Similarly, by considering the higher derivatives of (\ref{cestbien}) with respect to $\alpha$, we conclude:
\begin{equation}
P_1=P_2=\cdots=0
\end{equation}

Let us briefly sketch how this is done. Assume that we have a non-zero $P_k$. Then we would have

\begin{eqnarray}
&&\frac{\partial^k}{\partial\alpha^k} \int_{A_{cut}}^{\infty} \frac{P_k}{A^k} e^{-\alpha A}dA \nonumber\\
&&=(-1)^k\int_{A_{cut}}^{\infty} P_k e^{-\alpha A}dA\label{proof}\\
&&=(-1)^k\frac{P_k}{\alpha} e^{-\alpha A_{cut}}=(-1)^k\frac{P_k}{\alpha}+O(1)+O(\alpha)+\cdots\label{proof1}
\end{eqnarray}

Integrating (\ref{proof1}) by $\alpha$, $k$ times, we have:
\begin{equation}
\int_{A_{cut}}^{\infty} \frac{P_k}{A^k} e^{-\alpha A}dA=(-1)^k \frac{P_k}{(k-1)!} \alpha^{k-1}\ln \alpha +\cdots
\end{equation}
As the term proportional to $\alpha^{k-1} \ln \alpha$ is absent in (\ref{PAE-AA}), we obtain $P_k=0$ for $k>0$.

Plugging these values and (\ref{implication}) to (\ref{Laurent}), we conclude:
\begin{equation}
\lim_{A\rightarrow \infty} P(A)=\frac{1}{<A>}\label{PA}
\end{equation}
\begin{equation}
\lim_{A\rightarrow \infty} N(A)=\frac{1}{<A>}e^{A/4}\label{naive}
\end{equation}

However, this $N(A)$ is a naive one without the projection constraint or the $SU(2)$ invariant subspace constraint. We will correct this in the next section

Now, let's plug (\ref{PA}) to the formula (\ref{compare}) and equate it with (\ref{crucial}). By matching order by order, we obtain followings:
\begin{equation}
\frac{<A^2>}{2<A>^2}-1=\lim_{A_{cut}\rightarrow \infty}\{-\frac{A_{cut}}{<A>}+\sum_{A<A_{cut}} N(A) e^{-A/4}\}\label{infinite0}
\end{equation}
\begin{equation}
-\frac{1}{3}\frac{<A^3>}{<A>^2}+\frac{1}{4}\frac{<A^2>^2}{<A>^3}=\lim_{A_{cut}\rightarrow \infty}\{\frac{A_{cut}^2}{2<A>}-\sum_{A<A_{cut}} N(A) e^{-A/4}A\} \label{infinite1}
\end{equation}
and so on. In other words, we can obtain the value for the following formula
\begin{equation}
\lim_{A_{cut}\rightarrow \infty} \{-\frac{A_{cut}^{n+1}}{(n+1)<A>}+\sum_{A<A_{cut}} N(A) e^{-A/4} A^n\}\label{limit}
\end{equation}
which is convergent. In other words, (\ref{infinite0}) and (\ref{infinite1}) are the cases when $n=0,1$ in the above formula. 

Given this, I want to note that $P(A)$($\equiv N(A)\exp(-A/4)$) is zero for most of the values, as it would be a big coincidence if a given random $A$ is a sum of the area eigenvalues. In other words, $P(A)$ is non-zero only for the set which is measure zero. We can fix this by introducing $P'(A)$ as the ``continutization'' of $P(A)$ as follows:
\begin{equation}
\frac{<A^2>}{2<A>^2}-1=\lim_{A_{cut}\rightarrow \infty}\{-\frac{A_{cut}}{<A>}+\int_0^{A_{cut}} P'(A)dA\}\label{infinite0prime}
\end{equation}
\begin{equation}
-\frac{1}{3}\frac{<A^3>}{<A>^2}+\frac{1}{4}\frac{<A^2>^2}{<A>^3}=\lim_{A_{cut}\rightarrow \infty}\{\frac{A_{cut}^2}{2<A>}-\int_0^{A_{cut}} P'(A)A dA\} \label{infinite1prime}
\end{equation}
and so on. In other words, we have certain non-diverging values for the following formula
\begin{equation}
\lim_{A_{cut}\rightarrow \infty} \{-\frac{A_{cut}^{n+1}}{(n+1)<A>}+\int_0^{A_{cut}} P'(A) A^n dA\}\label{limitprime}
\end{equation}

Furthermore, even though it may sound redundant, we want to note that this convergence implies that $P_n=0$ for $n>0$. To see this, let's consider a non-zero $P_n$. Then, for large $A$ we have the following:
\begin{equation}
P'(A) A^n=(\frac{1}{<A>}+\frac{P_n}{A^n})A^n=\frac{A^n}{<A>}+P_n
\end{equation}

When the above term is plugged into the right-hand side of (\ref{limitprime}) and integrated, the potential divergence of the term proportional to $A_{cut}^{n+1}$ is removed, but the integration of $P_n$ survives, which yields roughly $P_n A_{cut}$, which is divergent in the limit $A_{cut}$ goes to infinity.

At this point, it may seem odd that $P'(A)$ doesn't receive any Laurent series corrections, but nevertheless still some corrections so that the left-hand sides of (\ref{infinite0prime}) and (\ref{infinite1prime}) are not zero. One may wonder such a function exists at all. An example of such function is following.

\begin{equation}
P'(A)=\frac{1}{<A>}+e^{-\beta A}
\end{equation}
for some positive $\beta$. We can clearly see that the above expression cannot be written in terms of Laurent series expansion exact in the limit in which $A$ is large, but that it is clearly different from $\frac{1}{<A>}$. Of course, this function does not satisfy (\ref{infinite0prime}) and (\ref{infinite1prime}), but one can guess that a suitable form for $P'(A)$ should be something of this kind. In the next section, we obtain an explicit solution for $P'(A)$

\section {Solution}
We suggest the following:
\begin{equation}
P'(A)=\frac{1}{<A>}+B(A)e^{-A}
\end{equation}
where $B(A)$ is a suitable polynomial.

Plugging this to (\ref{limitprime}), we obtain:
\begin{equation}
\int_0^{\infty}B(x)x^n e^{-x}dx= C_n
\end{equation}
for a suitable $C_n$. For example, from (\ref{infinite0prime}) and from (\ref{infinite1prime}), we obtain:
\begin{equation}
C_0=\frac{<A^2>}{2<A>^2}-1
\end{equation}
\begin{equation}
C_1=\frac{1}{3}\frac{<A^3>}{<A>^2}-\frac{1}{4}\frac{<A^2>^2}{<A>^3}
\end{equation}

Now recall Laguerre polynomial:
\begin{equation}
L_n(x)=\sum_{k=0}^n \frac{(-1)^k}{k!} {n\choose k} x^k
\end{equation}
Then, we have:
\begin{equation}
\int_0^{\infty}B(x)L_n(x) e^{-x}dx= \sum_{k=0}^n \frac{(-1)^k}{k!} {n\choose k} C_k
\end{equation}
Since
\begin{equation}
\int_0^{\infty}e^{-x} L_m(x) L_n(x) dx=\delta_{mn}
\end{equation}

We have:
\begin{equation}
B(x)=\sum_{n=0}^{\infty}d_n L_n(x)
\end{equation}
where
\begin{equation}
d_n=\sum_{k=0}^n \frac{(-1)^k}{k!} {n\choose k} C_k
\end{equation}

Therefore, the solution is:
\begin{equation}
P'(A)=\frac{1}{<A>}+\sum_{n=0}^{\infty}d_n L_n(A)e^{-A}
\end{equation}
\section {Corrections}
In case of the $U(1)$ symmetry reduced framework, if we consider the extra condition, the so-called ``projection constraint'' $\sum_i m_i=0$ \cite{Meissner}, the black hole degeneracy (\ref{naive}) will be reduced. We will calculate the reduced black hole degeneracy by multiplying the probability that this condition is satisfied to our earlier naive black hole degeneracy. Before doing so, let us explain what $m_i$s are. $m_i$ are half integers which satisfy:

\begin{equation}
-j_i\leq m_i \leq j_i
\end{equation}
where $j_i$s are given in section 3. In other words, it has the same structure as 3d-angular momentum in quantum mechanics.

Given this, let's define $x\equiv\sum_i m_i$. Then, we have:
\begin{equation}
\Delta x^2=\sum_i \Delta m_i^2\label{x}
\end{equation}

Of course, we can express $\Delta m_i^2$ as in terms of $j_i$, as

\begin{equation}
\Delta m_i^2=(\sum_{m_i=-j_i}^{j_i} m_i^2)/(2 j_i +1)
\end{equation}

Then, (\ref{x}) becomes
\begin{equation}
\Delta x^2=\sum_i \Delta m_i^2(j_i) \label{xlI}
\end{equation}

Now, we need to calculate $F_{j_i}$, the number of times given $j_i$ appears on the right-hand side of the above equation. From thermodynamics consideration or observation from Domagala-Lewandowski-Meissner trick \cite{Meissner, Domagala}, it is obvious that this frequency is proportional to $e^{-A_{j_i}/4}$, where we remind the reader that $A_{j_i}$ is given by:

\begin{equation}
A_{j_i}=8\pi\gamma \sqrt{j_i(j_i+1)}
\end{equation}
in isolated horizon case.
Also, taking into account the fact that the total sum of area of each segment in the black hole horizon is $A$, we obtain:

\begin{eqnarray}
A=\sum_j A_j(\frac{A}{<A>} e^{-A_j/4}) =\sum A_j F_j\nonumber\\
F_j=\frac{A}{<A>} e^{-A_j/4}
\end{eqnarray}

For a macroscopic black hole, (\ref{xlI}) can be written as:

\begin{equation}
\Delta x^2=\sum_j F_j \Delta m^2(j)=A C
\end{equation}
where $C$ is an unimportant constant which one can calculate from the area spectrum and $\Delta m^2(j)$.

Now, noticing that the distribution of $x$ reaches Gaussian for macroscopic black hole by the well-known theorem in statistics, we can write $p(0)$, the probability that $x=0$ as follows:

\begin{equation}
p(0)\approx\int_{x=-1/4}^{x=1/4}\frac{1}{\sqrt{2 \pi C A}} e^{-x^2/(2C A)}\approx\frac{1}{2\sqrt{2 \pi C A}}
\end{equation}

Therefore, the correct degeneracy is given by:

\begin{equation}
N_{cor}(A)=\frac{1}{2<A>\sqrt{2 \pi C A}}e^{A/4}=\frac{1}{D\sqrt{A}}e^{A/4}
\end{equation}
where D is an unimportant constant. (Remember that $<A>$ is merely a constant which one can calculate from the area spectrum and which doesn't depend on the black hole area $A$.)
Therefore, we conclude that the black hole entropy is given by:

\begin{equation}
S=\ln N_{cor}(A)=\frac{A}{4} -\frac{1}{2} \ln A+O(1)
\end{equation}

The logarithmic corrections to the Bekenstein-Hawking entropy in case of the fully $SU(2)$ framework can be obtained similarly. For a detailed discussion, please read \cite{LivineTerno}. See also, \cite{ENP1, ENP2, ENP3}.
\section {Discussions and Conclusions}
In this paper, we have obtained the infinite relations between the area spectrum and the naive black hole degeneracy, and obtained an explicit solution for them. This could be especially important for the analysis of mini-black holes that have a possibility to be created at LHC, since this is the case when $A$ is small and the deviation of black hole entropy from the Bekenstein-Hawking entropy is not negligible. It is suggested in \cite{ToAppear} how one can check the black hole entropy by measuring Hawking radiation spectrum.

\section{Acknowledgements}
We thank Jong-Hyun Baek for the helpful and crucial discussions. This work was supported by the National Research Foundation of Korea (NRF) 
grants 2012R1A1B3001085 and 2012R1A2A2A02046739.


\begin{thebibliography}{9}

\bibitem{Bekenstein}
  J.~D.~Bekenstein,
  ``Black holes and entropy,''
  Phys.\ Rev.\  D {\bf 7}, 2333 (1973).

\bibitem{Hawking}
  S.~W.~Hawking,
  ``Particle Creation By Black Holes,''
  Commun.\ Math.\ Phys.\  {\bf 43}, 199 (1975)
  [Erratum-ibid.\  {\bf 46}, 206 (1976)].

\bibitem{corrections}
R.~K.~Kaul and P.~Majumdar,
  ``Logarithmic correction to the Bekenstein-Hawking entropy,''  Phys.\ Rev.\ Lett.\  {\bf 84}, 5255 (2000)  [gr-qc/0002040].  

A.~Ghosh and P.~Mitra,
  ``A Bound on the log correction to the black hole area law,''  Phys.\ Rev.\ D {\bf 71}, 027502 (2005)  [gr-qc/0401070].  

A.~Ghosh and P.~Mitra,
  ``An Improved lower bound on black hole entropy in the quantum geometry approach,''  Phys.\ Lett.\ B {\bf 616}, 114 (2005)  [gr-qc/0411035].  

A.~Corichi, J.~Diaz-Polo and E.~Fernandez-Borja,
  ``Quantum geometry and microscopic black hole entropy,''  Class.\ Quant.\ Grav.\  {\bf 24}, 243 (2007)  [gr-qc/0605014].  

H.~Sahlmann,
  ``Entropy calculation for a toy black hole,''  Class.\ Quant.\ Grav.\  {\bf 25}, 055004 (2008)  [arXiv:0709.0076 [gr-qc]].  

I.~Agullo, G.~J.~Fernando Barbero, E.~F.~Borja, J.~Diaz-Polo and E.~J.~S.~Villasenor,
  ``The Combinatorics of the SU(2) black hole entropy in loop quantum gravity,''  Phys.\ Rev.\ D {\bf 80}, 084006 (2009)  [arXiv:0906.4529 [gr-qc]].  

\bibitem{Meissner}
  K.~A.~Meissner,
  ``Black hole entropy in loop quantum gravity,''
  Class.\ Quant.\ Grav.\  {\bf 21}, 5245-5252 (2004).
  [gr-qc/0407052].

\bibitem{Rovelli}
  C.~Rovelli,
  ``Black hole entropy from loop quantum gravity,''
  Phys.\ Rev.\ Lett.\  {\bf 77}, 3288-3291 (1996).
  [gr-qc/9603063].

\bibitem{Discreteness}
  C.~Rovelli, L.~Smolin,
  ``Discreteness of area and volume in quantum gravity,''
  Nucl.\ Phys.\  {\bf B442}, 593-622 (1995).
  [gr-qc/9411005].

\bibitem{General}
  S.~Frittelli, L.~Lehner, C.~Rovelli,
  ``The Complete spectrum of the area from recoupling theory in loop quantum gravity,''
  Class.\ Quant.\ Grav.\  {\bf 13}, 2921-2932 (1996).
  [gr-qc/9608043].

\bibitem{Ashtekar}
  A.~Ashtekar, J.~Lewandowski,
  ``Quantum theory of geometry. 1: Area operators,''
  Class.\ Quant.\ Grav.\  {\bf 14}, A55-A82 (1997).
  [gr-qc/9602046].

\bibitem{Domagala}
  M.~Domagala, J.~Lewandowski,
  ``Black hole entropy from quantum geometry,''
  Class.\ Quant.\ Grav.\  {\bf 21}, 5233-5244 (2004).
  [gr-qc/0407051].

\bibitem{Tanaka}
  T.~Tanaka and T.~Tamaki,
  ``Black hole entropy for the general area spectrum,''
  arXiv:0808.4056 [hep-th].

\bibitem{isolated horizon}
  A.~Ashtekar, J.~C.~Baez and K.~Krasnov,
  ``Quantum geometry of isolated horizons and black hole entropy,''  Adv.\ Theor.\ Math.\ Phys.\  {\bf 4}, 1 (2000)  [gr-qc/0005126].  

\bibitem{Compositions}
``Digital Library of Mathematical Functions. 2011-08-29. National Institute of Standards and Technology from http://dlmf.nist.gov''
Chapter 26.11 Integer Partitions: Compositions, http://dlmf.nist.gov/26.11

\bibitem{MacMahon}
P.A. MacMahon, Combinatory analysis, volume I, page 150, 151, 154. Cambridge University Press, 1915-16.

\bibitem{LivineTerno} 
  E.~R.~Livine and D.~R.~Terno,
  ``Quantum black holes: Entropy and entanglement on the horizon,''
  Nucl.\ Phys.\ B {\bf 741}, 131 (2006)
  [gr-qc/0508085].
\bibitem{ENP1}
  J.~Engle, A.~Perez and K.~Noui,
  ``Black hole entropy and SU(2) Chern-Simons theory,''  Phys.\ Rev.\ Lett.\  {\bf 105}, 031302 (2010)  [arXiv:0905.3168 [gr-qc]].  

\bibitem{ENP2}
  J.~Engle, K.~Noui, A.~Perez and D.~Pranzetti,
  ``Black hole entropy from an SU(2)-invariant formulation of Type I isolated horizons,''  Phys.\ Rev.\ D {\bf 82}, 044050 (2010)  [arXiv:1006.0634 [gr-qc]].  

\bibitem{ENP3}
  J.~Engle, K.~Noui, A.~Perez and D.~Pranzetti,
  ``The SU(2) Black Hole entropy revisited,''  JHEP {\bf 1105}, 016 (2011)  [arXiv:1103.2723 [gr-qc]].  
\bibitem{ToAppear}
  Y.~Yoon,  ``Quantum corrections to Hawking radiation spectrum,''
to appear.
\end{thebibliography}
\end{document}